\documentclass[aps,prd,amssymb,amsmath,nofootinbib,showpacs,12pt]{revtex4}
\usepackage{graphicx}
\usepackage{dcolumn}
\usepackage{bm}

\usepackage{amssymb}
\usepackage{amsmath}
\usepackage{hyperref}

\def\d{\rm d}

\begin{document}
\title{Self-force on a scalar charge in radial infall from rest using the Hadamard-WKB expansion}
\author{Paul R. Anderson}
 \email{anderson@wfu.edu}
 \affiliation{Department of Physics, Wake Forest University, Winston-Salem, NC  27109, USA}
\author{Ardeshir Eftekharzadeh}
 \email{eftekhar@wam.umd.edu}
\author{B. L. Hu}
 \email{hub@physics.umd.edu}
\affiliation{Department of Physics, University of Maryland,
College Park, MD 20742, USA}
\date{\small September 29, 2005}

\begin{abstract}
We present an analytic method based on the Hadamard-WKB expansion to
calculate the self-force for a particle with scalar charge that
undergoes radial infall in a Schwarzschild spacetime after being
held at rest until a time $t = 0$.  Our result is valid in the case
of short duration from the start. It is possible to use the
Hadamard-WKB expansion in this case because the value of the
integral of the retarded Green's function over the particle's entire
past trajectory can be expressed in terms of two integrals over the
time period that the particle has been falling. This analytic result
is expected to be useful as a check for numerical prescriptions
including those involving mode sum regularization and for any other
analytical approximations to self-force calculations.
\end{abstract}
\pacs{04.25.Nx, 04.30.Db, 04.70.Bw}

\maketitle

\section{Introduction}

{\it This latest version has a new appendix C containing the Erratum
in Ref. [41] to the version published in Physical Review D [40]. In this version we
have made the corrections indicated therein. }

Computation of the self-force for a point particle in orbit around a
black hole is a topic of active research today
\cite{selforce,Quinn,PoiDet} prompted by the preparation of
gravitational wave detectors such as LISA which are capable of
detecting gravitational waves emitted when a compact object falls
into a supermassive black hole~\cite{lisa1,lisa2}.

An exact expression for the self-force in a black hole spacetime
has been obtained only for the cases of a scalar or electric
charge held at rest in a Schwarzschild spacetime
\cite{SW,Lohiya,Wis,Ros} and an electric charge held at rest on
the symmetry axis of a Kerr spacetime~\cite{LL,Piazzese}.
Approximate analytical expressions have been obtained using
Green's functions for scalar, electric, and gravitational charges
in various weak field limits~\cite{DeW2,Gal,PfePoi}.  Most other
calculations have involved the use of mode sum
techniques~\cite{BarOri,BMNOS} in cases of high symmetry such as a
static charge~\cite{burko,burko-liu,burko-liu-2}, radial
infall~\cite{BB,BL}, a circular
orbit~\cite{BBc,DMW,diaz-rivera-et-al,hikida-et-al}, or a slightly
elliptical orbit~\cite{diaz-rivera-et-al}.

Since the mode-sum regularization procedure has been developed
extensively, aiming at practical calculations, it is important to
find independent ways to check its accuracy and reliability.  This
of course also applies to any other method, numerical or analytical
that might be developed in the future.  To date the checks on
various mode-sum regularization procedures which we are aware of
fall into three categories:  1) comparisons with exact analytical
results, 2) comparisons with analytical approximations, 3)
comparisons of the results of one mode-sum regularization technique
with those of another.  Most of the comparisons that have been made
fall into the third
category~\cite{detweiler-poisson,hikida-et-al,diaz-rivera-et-al,DMW,BL,
lousto}.  Comparisons with exact analytical results are of course
the most reliable but also the most limited. They have only been
done in the case of static scalar and electric charges held at rest
in a Schwarzschild spacetime~\cite{burko}.  Comparison with an
analytic approximation in the weak field limit has been made for a
static scalar charge at rest in an axisymmetric
spacetime~\cite{burko-liu-2}. In this paper we present results which
can be used as a check in the second category.  However, unlike most
previous analytical approximations to the self-force, ours is valid
in the strong field as well as the weak field limit. Specifically,
we consider the case of a particle with scalar charge which is held
at rest until a time $t=0$ and subsequently falls radially towards a
Schwarzschild black hole.

In a previous paper \cite{AndHu} (which we will refer to as Paper
I) we introduced an approximation for the computation of the
self-force of a scalar charge in a Schwarzschild spacetime using
the Hadamard-WKB expansion \cite{Sch,DeW,Chr,ALN,Wald,AHS,PhiHu}
which can account for the radiation reaction effects from the
charge's recent past history. This is an expansion for the tail
term part of the retarded Green's function and is valid in a
spacetime region close to the charge. A similar approach has been
taken to compute part of the gravitational self-force in an
arbitrary spacetime by Anderson, Flanagan,and Ottewill~\cite{afo}.
In Paper I an eighth order WKB expansion was used to compute the
nonvanishing part of the tail term to order $(x-x')^6$.  Since
then we have obtained results to order $(x-x')^{14}$ using a 16th
order WKB expansion.

Recently Anderson and Wiseman~\cite{AndWis} investigated the
convergence of the Hadamard-WKB expansion for the self-force. They
posited that convergence is obtained as long as the point separation
in proper time is small enough so that the null geodesics emanating
from the point at the earlier time have not had time to re-intersect
the particle's path. We know the absence of caustics is perhaps the
least stringent criterion for the validity of a quasilocal expansion
(e.g., the domain of validity of Riemann normal coordinates). After
computing this proper time separation for the case of a particle
orbiting a black hole in a circular path, they use the result of
Paper I [Hadamard-WKB expansion to  order $(x-x')^6$] to evaluate
the convergence of the series expansion for the self-force.

Anderson and Wiseman go on to show that the primary constituent of
the self-force comes from the tail term of the Green's function when
the points are too widely separated for this expansion to converge.
This would seem to rule out the possibility of using the
Hadamard-WKB expansion by itself to compute the self-force, which is
not surprising, as the quasilocal expansion has this intrinsic
limitation to begin with.

However, we want to point out that there is at least one case in
which it is possible to determine the entire self-force using the
Hadamard-WKB expansion.  If the particle is held at rest until a
time $t = 0$ after which it falls radially towards the black hole,
then the entire self-force can be determined using the Hadamard-WKB
expansion so long as the particle has not fallen too far from its
starting point.  While the range over which the particle falls must
be small, the results can serve as a useful check on other methods
of computing the self-force.

In Section \ref{sec:review} a brief review is given of the general method
we use to compute the self-force.  In Section
\ref{sec:computation} the self-force is computed for
the case in which a particle is held at rest until time $t = 0$ and
subsequently falls towards the event horizon of a Schwarzschild
black hole.  We derive a closed expression for the self-force in
terms of integrals over the tail part of the retarded Green's
function when the points are split by only a small amount. The
splitting is between the point where the particle is at a time $t
>0$ and various other points inside the past lightcone of this
point. We then use the Hadamard-WKB expansion described in Paper I
to derive power series expressions for the nonzero components of the
self-force.  In Section \ref{sec:example} the series are evaluated
for a specific example. The results are given in Table I and Figure
I. Some details of the calculations are given in Appendices A and B
and the coefficients derived for the series are also given in
Appendix B.

\section{Review of Method}
\label{sec:review}

Consider a massless scalar field with a source term consisting of
a point particle of scalar charge $q$.  The wave equation
is~\cite{Wis,Quinn}\footnote{Note that our conventions differ from those of
\cite{Wis,Quinn}.  One can obtain those of Ref.~\cite{Quinn} by letting
$\Phi \rightarrow \Phi/\sqrt{4 \pi}$, $q \rightarrow \sqrt{4 \pi} q$, and
$G_R \rightarrow G_R/(4 \pi)$.  Those of Ref.~\cite{Wis} can be obtained by
letting $\Phi \rightarrow -\Phi/\sqrt{4 \pi}$ and $q \rightarrow \sqrt{4 \pi} q$.}
\begin{eqnarray}
\Box \Phi(x) &=& - \rho(x)   \label{eq:boxphi} \\
\rho(x) &=& q \int_{-\infty}^\infty
\frac{\delta^4\big(\,x,z(\tau)\,\big)}{\sqrt{-g}} d \tau \nonumber
\end{eqnarray}
where $\tau$ is the proper time and $z(\tau)$ is the trajectory
of the particle. A formal solution to this equation can be
obtained using the retarded Green's function:
\begin{equation}
\Phi(x) = q \int_{-\infty}^{\infty} G_R\big(x,z(\tau')\big) d\tau'
\;. \label{eq:phi}
\end{equation}
 The self-force is given formally by
\begin{equation} f_\mu(\tau) = q \left[\nabla_\mu \Phi(x) \right]_{x = z(\tau)
} \;.
\end{equation}
This expression is divergent and must be regularized.
Quinn~\cite{Quinn} has shown that the regularized expression for the
self-force can be split into a local term plus a finite integral
over the gradient of the retarded Green's function.  The latter is
often called the ``tail term''.

The Hadamard expansion for the retarded Green's function
is~\cite{Quinn,PhiHu,ALN,Wald}\footnote{In Eq. (1.1) of Paper I this equation was written with an incorrect factor of $\sigma(x,x')$ in the denominator of the first term.  The relationships between our definitions of $u$ and $v$ and those of Ref. [2] were also given incorrectly.  Taking into account the previously mentioned difference in the definition of the retarded Green's function, the correct relationship is that the definition of $u$ in Eq. (4) is equivalent to that in Ref. [2] while the definition of $v$ differs from that of Ref. [2] by a factor of $1/2$.  No other equations in Paper I are affected and the results remain unchanged.}
\begin{eqnarray}
 G_R(x,x') &=& \theta(x,x') \left\{ \frac{u(x,x')}{4 \pi } \delta[\sigma(x,x')]
 - \frac{v(x,x')}{8 \pi} \theta[-\sigma(x,x')] \right\} \;.
\label{eq:GDS}
\end{eqnarray}
Here $\theta[-\sigma(x,x')]$ is defined to be one if the point $x'$
is inside the light cone of the point $x$ and zero otherwise, while
$\theta(x,x')$ is defined to be one if the point $x$ resides in the
future of a space-like hypersurface involving the point $x'$ and
zero otherwise.  The quantity $\sigma(x,x')$ is equal to one-half
the square of the proper distance between $x$ and $x'$ along the
shortest geodesic connecting them.  The function $v(x,x')$
contributes to the tail part of the self-force~\cite{Quinn}.  It
obeys the equation
\begin{equation} \Box_x v(x,x') = 0  \end{equation} and is symmetric under
the exchange of the two points,
i.e. $v(x,x') = v(x',x)$. In a general spacetime~\cite{DeW,Chr}
\begin{equation}
v(x,x) = - \frac{1}{6} R(x) \;.
\end{equation}
The tail term for the self-force obtained from Eqs.
(\ref{eq:boxphi}) and (\ref{eq:phi}) can be written in the form
(see, e.g., ~\cite{AndWis})
\begin{eqnarray}
\left(f_\mu (\tau) \right)_{{\rm tail}} &=& -\frac{q^2}{8\pi}
 \int_{\tau_0}^\tau \left( \frac{\partial}{\partial x^\mu} v[x,z(\tau')]\right)_{x=z(\tau)} d \tau' \nonumber \\
&  & + \, q^2 \int_{-\infty}^{\tau_0} \left(
\frac{\partial}{\partial x^\mu} G_R[x,z(\tau')] \right)_{x=z(\tau)}
\, d\tau' \;.  \label{eq:selfforce}
\end{eqnarray}
It is necessary that $\tau_0$ be chosen so that the Hadamard-WKB
expansion for $v$ is valid throughout the region of integration of
the first integral in Eq.~(\ref{eq:selfforce}).

An expansion for $v(x,x')$  in Schwarzschild spacetime was found in
Paper I using a WKB expansion for the Euclidean Green's function.
The WKB expansion is obtained via an iteration procedure and is
increased by two orders upon each iteration.  From a WKB expansion
of order $(2N)$ one can obtain an expansion for $v(x,x')$ that
includes terms up to order $(x-x')^{2N-2}$.  For the metric
\begin{equation}
ds^2 = -(1-2M/r) dt^2 + dr^2/(1-2M/r) + r^2
d\Omega^2 \;, \label{eq:metric}
\end{equation}
the expansion for $v$ is of the form
\begin{eqnarray}
v(x,x') &=& \sum_{i,j,k=0}^\infty v_{ijk}(r) (t-t')^{2i} (\cos \gamma -
1)^j (r-r')^k
 \label{eq:vexp}
\end{eqnarray}
with
\begin{equation} \cos \gamma \equiv \cos \theta \cos \theta' +
\sin \theta \sin \theta' \cos (\phi - \phi') \;. \label{eq:cosgamma}
\end{equation}

We have recently found that it is possible to extend this expansion
to order $(x-x')^{2N-1}$ by using the fact that $v(x,x')$ is a
symmetric function.  If one takes $(2N-1)$ partial derivatives of
$v(x,x')$ with respect to some combination of the coordinates $x^a$
and  sets $x' = x$ then by symmetry it must be true that taking the
same combination of derivatives with respect to $x'^a$ and setting
$x' = x$ results in an equivalent expression. Using the expansion
(\ref{eq:vexp}) results in the coefficients of terms of odd powers
of $(x-x')$ being expressed in terms of the coefficients of terms of
smaller even powers.  The coefficients of  terms of even powers of
$(x-x')$ cannot be obtained in this way and must be obtained from
the WKB expansion.

\section{Computation of the self-force}
\label{sec:computation}

We now proceed to compute the self-force for a particle of mass $m$
and scalar charge $q$ falling radially from rest.  We assume that
the particle has been held fixed at $r=r_0$, $\theta = \theta_0$,
and $\phi = \phi_0 $ from $t=-\infty$ to $t=0$ which corresponds to
proper time $\tau = 0$. At time $t = 0$ the particle is released,
with no initial velocity, and subsequently falls radially towards a
Schwarzschild black hole of mass $M$.  We shall denote the spacetime
point at which it is released by $y_0$.  Our objective is to
calculate the self-force exerted on the particle at a proper time
$\tau >0$ when the particle is at the point $y = (t, r, \theta_0,
\phi_0)$.

 The first term in Eq.~(\ref{eq:selfforce}) can be computed
using the Hadamard-WKB expansion.  The results of this computation
are discussed below. The second term is more difficult.  We next
show that it too can be computed using the Hadamard-WKB expansion
due to the fact that the charge is stationary until it begins
falling at time $t=0$.

First consider the problem of computing the field $\Phi_{static}$ at
the point $y$ due to a completely static charge $q$ at the position
$r=r_0, \theta = \theta_0, \phi = \phi_0$. The solution to this
problem in a Schwarzschild spacetime has been given by
Wiseman~\cite{Wis}. For the above locations of the charge and field
point, with a metric of the form (\ref{eq:metric}), and taking
into account a difference in conventions, it can be written as
\begin{eqnarray}
\Phi_{static}(y) = \frac{1}{4\pi}q \sqrt{1 - \frac{2M}{r_0}} \,
\frac{1}{r_0 - r} \;. \label{eq:wiseman}
\end{eqnarray}

 On the other hand one can use Eq.~(\ref{eq:phi}), to write
the field due to the static charge as
\begin{eqnarray} \Phi_{static}(y) = q
\int_{-\infty}^{\infty} G_R[t,r;t'(\tau'),r_0] d\tau' \;.
\end{eqnarray}
Here and for the rest of this section we suppress the dependence of
various quantities on the angles $\theta_0$ and $\phi_0$.  Dividing
the integration region in the same way as was done in
Eq.~(\ref{eq:selfforce}) and using the Hadamard expansion
(\ref{eq:GDS}) one can write this latter equation as
\begin{eqnarray}
\Phi_{static}(y) &=&  \frac{q}{4 \pi}\sqrt{1-\frac{2M}{r_0}}
\int_{0}^{t} u[t,r;t',r_0]\,
\delta[\sigma(t,r;t',r_0)]\,  dt' \nonumber \\
&-&\frac{q}{8 \pi} \sqrt{1 - \frac{2M}{r_0}} \int_{0}^{t_R}
v[t,r;t',r_0] \,
 dt' \; +q \int_{-\infty}^{0}
G_R[t,r;t'(\tau'),r_0] \, d\tau' \label{eq:phitest}
\end{eqnarray}
In the first two terms the integration variable has been changed
from $\tau'$ to $t'$.  The two theta functions in Eq.~(\ref{eq:GDS})
result in an upper limit for the second integral which is equal to
the retarded time $t_R$ which is given in Eq.\ (\ref{eq:tR}). The
time $t$ is taken to be the time that it would take a particle to
fall from $r_0$ to $r$ assuming that it starts at rest.  Then the
third term on the right in Eq.\ (\ref{eq:phitest}) is the same,
except for the gradient and a factor of $q$, as the second term on
the right in Eq.~(\ref{eq:selfforce}). The value of this term can be
obtained by computing the other three terms in the equation.  The
term on the left is given in Eq.~(\ref{eq:wiseman}) and the second
term on the right can be computed using the Hadamard-WKB expansion.

To calculate the first term on the right in Eq.~(\ref{eq:phitest}),
we note that the argument of the delta function vanishes on the
light cone of the point $y$. Since the charge is static
\begin{eqnarray}
\sigma(t,r;t',r_0) &=& \sigma(t,r;t_R,r_0) + \left(
\frac{\partial}{\partial t'}
      \sigma(t,r;t',r_0) \right)|_{t'=t_R} \; (t'-t_R) + ... \nonumber \\
&=& \sigma_{t_{R}} (t'-t_{R}) + ... \;, \label{eq:sigma}
\end{eqnarray} where the shorthand notation $\sigma_\mu \equiv
\sigma_{; \mu}$ has been used. Then
\begin{equation}
\delta(\sigma(t,r;t_R,r_0)) = \delta[\sigma_{t_R} (t'-t_R)] =
\frac{\delta(t'-t_R)}{|\sigma_{t_R} |} \;.
\end{equation}

Next one must calculate $u(y,y_R)$ with $y$ and $y_R \equiv
(t_R,r_0,\theta_0,\phi_0)$ connected by a null radial geodesic. By
substituting the Hadamard expansion into the equation satisfied by
the Green's function it is possible to show that in
general~\cite{Sch,DeW,Chr}
\begin{eqnarray}
\label{eq:defu}  u(x,x') &=& \Delta^{1/2}(x,x') \\
\Delta(x,x') &=& -\frac{{\rm det} (-\sigma_{;\mu
\nu'})}{\sqrt{-g(x)} \sqrt{-g(x')}} \nonumber
\end{eqnarray}
Thus what remains is to calculate $\sigma_{t_R}$ and $\sigma_{;\mu
\nu'}$  for the two points $x = y$ and $x' = y_R$.  Although there
may be some simple way to reason out the answer, as shown in
Appendix A, it can be obtained by solving the geodesic equations and
integrating the result to obtain the proper distance along the
geodesic. The result is
\begin{eqnarray}
\sigma_{t_R}(t,r;t_R,r_0) &=& r_0 - r  \nonumber \\
 u(t,r;t_R,r_0) &=& 1  \;. \label{eq:u}
\end{eqnarray}

Substituting Eq.~(\ref{eq:u}) into Eq.~(\ref{eq:phitest}) and
computing the integral one finds that
\begin{equation}
\int_{-\infty}^{0} G_R[t,r;t'(\tau'),r_0] \, d\tau' = \frac{1}{8
\pi} \sqrt{1 - \frac{2M}{r_0}} \int_{0}^{t_R} v[t,r;t',r_0] dt'
\label{eq:GR}
\end{equation}
With the definitions
\begin{subequations}
\begin{eqnarray}
\label{eq:phis1} \Phi_s(y) &=& \frac{q}{8 \pi} \sqrt{1 -
\frac{2M}{r_0}} \int_{0}^{t_R} v[t,r;t',r_0] \,  \;\;dt'
\\\nonumber
\\  \label{eq:phif}\Phi_f (y) &=& \frac{q}{8\pi} \int_{0}^{\tau}
v[t,r;t'(\tau'),r'(\tau')] \, d \tau'
\end{eqnarray}
\end{subequations}
 Eq.~(\ref{eq:selfforce}) becomes
\begin{eqnarray}
\label{eq:selfforce1}f_\mu(\tau) = q \left[\frac{\partial}{\partial
y^\mu} (\Phi_s(y)-\Phi_f(y) )\right] \;.
\end{eqnarray}
Here the facts that in a Schwarzschild spacetime $v(x,x) = 0$ and
(as shown by our expansion) $v(y,y_R) = 0$, have been
used to interchange the order of integration and differentiation.
Note that this derivation only works for the
time and radial components of the self-force. Because of spherical
symmetry, the angular components of the self-force  for a radial
trajectory are zero. Finally the subscript ``tail'' has been dropped
because for a geodesic trajectory the local part of the self-force
is zero in a Schwarzschild spacetime~\cite{Quinn}.

As a result of Eq. (\ref{eq:selfforce1}), the problem of calculating
the self-force reduces to calculating $\Phi_s$ and $\Phi_f$. This is
an exact result.  We now calculate the right hand side of Eq.
(\ref{eq:selfforce1}) using the Hadamard-WKB expansion for $v(x,x')$
whose form is given in Eq.\ (\ref{eq:vexp}). For radial geodesics,
$\cos \gamma = 1$, so only the coefficients $v_{i0k}(r)$ contribute.
The result for $\Phi_s$ is
\begin{eqnarray}
\label{eq:phis2} \Phi_s (y) &=& \frac{q}{8\pi}\sqrt{1 -
\frac{2M}{r_0}}\sum_{i,k=0}^\infty
\left(\frac{1}{2i+1}\right)v_{i0k}(r)\left[ t^{2i+1} - (t -
t_R)^{2i+1} \right] (r - r_0)^k
\end{eqnarray}
To calculate $\Phi_f$ one can use the geodesic equations
(\ref{eq:geodesic}) to convert the integral~(\ref{eq:phif}) to an
integral over the radial coordinate $r$.  One can further solve the
geodesic equations to obtain the trajectory $t(r)$.  After
substituting the Hadamard-WKB expansion for $v$, the
integral~(\ref{eq:phif}) can be computed numerically.

An alternative is to expand all relevant quantities in both $\Phi_s$
and $\Phi_f$ in Taylor series about $r_0$.  This allows one to
compute the integrals analytically order by order.  The derivation
is given in more detail in Appendix B.  We find
\begin{subequations}
\begin{eqnarray}
f_t(\tau) &=& \frac{q^2}{22400\,\pi \,
     r_0^2} \sqrt{1 - \frac{2M}{r_0}}
\left(40 - \frac{106M}{r_0} \right) \left(\frac{r_0 -
r}{r_0}\right)^3
     + O\left[\left(\frac{r_0-r}{r_0}\right)^4 \right] \\
f_{r*}(\tau) &=& - \frac{3 q^2}{11200 \pi r_0^2}\,
\sqrt{\frac{2M}{r_0}} \left(1 - \frac{2M}{r_0} \right)
\left(\frac{r_0-r}{r_0}\right)^{5/2} +
O\left[\left(\frac{r_0-r}{r_0} \right)^{7/2} \right]
\end{eqnarray}
\label{eq:fmu}
\end{subequations}
with $r*$ the Regge-Wheeler coordinate defined by
\begin{equation}
r^* \equiv r + 2M \log \left(\frac{r-2M}{2M} \right) \;.
\label{eq:rstar}
\end{equation}
 It turns out that each subsequent order of the WKB
expansion adds another term to the series. Using a 16th order WKB
expansion we have results for a total of six terms in the expansions
for both $f_t$ and $f_{r*}$.  The coefficients of these terms are
displayed in Appendix B.

\section{Specific Example}
\label{sec:example}

As a specific example we have chosen the case in which a particle
begins falling from rest at $r* = 40M$ at time $t = \tau = 0$, with
$r*$ defined in Eq.\ (\ref{eq:rstar}).  Using the series in Appendix
B, the $f_t$ and $f_r*$ components of the self-force have been
computed. The results are displayed in Table I and Figure I. The
relative errors in the table are estimated by taking the absolute
value of the ratio between the last term used in the series
expansion (\ref{eq:fmu}) for a given component of the self-force and
the entire series for that component.


{\center{
\begin{table}
 \begin{tabular}{|c|c|c|c|c|}
      \hline
  $ \; r/M \; $ & $10^{14} (M^2/q^2) f_t$  &  Error & $10^{14} (M^2/q^2) f_{r^*}$ & Error \\ \hline
      $34.25$ & $ 6.93066 $ &  $ 6.9 \times 10^{-6}$ & $ -5.7384 $ &  $  3.4 \times 10^{-5}$ \\ \hline
      $34.00$ & $ 120.72$ &  $ 4.4 \times 10^{-4}$ & $ -96.78$ &  $1.4 \times 10^{-3}$ \\ \hline
      $33.75$ & $ 616$ &  $ 3.4 \times 10^{-3}$ & $ -520$ &  $ 8.4 \times 10^{-3}$ \\ \hline
      $33.50$ & $ 2080$ &  $ 0.013$ & $ -1840$ &  $  0.25$ \\ \hline
      $33.25$ & $ 5680$ &  $ 0.031$ & $ -5200$ &  $ 0.052$ \\ \hline
      $33.00$ & $ 13800$ &  $ 0.059 $ & $ -13200$ &  $  0.089$ \\ \hline
      $32.75$ & $ 30000$ &  $ 0.10$ & $ -30000$ &  $  0.13    $ \\ \hline
      $32.50$ & $ 66000  $ &  $ 0.14     $ & $ -62000  $ & $  0.18    $ \\ \hline
  \end{tabular}
  \caption{The dependence of the temporal ($f_t$) and radial ($f_{r^*}$)
components of the self-force on the radial distance $r/M$ are given
for a particle undergoing radial infall after being held at rest
until a time $t=0$.  The particle's initial location is at $r^* =
40M$ which corresponds to $r \approx 34.43 M$.
  The error shown for each case is an estimate of the relative error and is obtained
  by taking the absolute value of the
  ratio of the last term used in the series for the self-force to the entire series.}
  \label{restab}
  \end{table}
}}

 {\center{
\begin{figure}
\begin{picture}(200,250)(-200,0)
\put(-250,0)
{\rotatebox{90}{\resizebox{8cm}{!}{\includegraphics{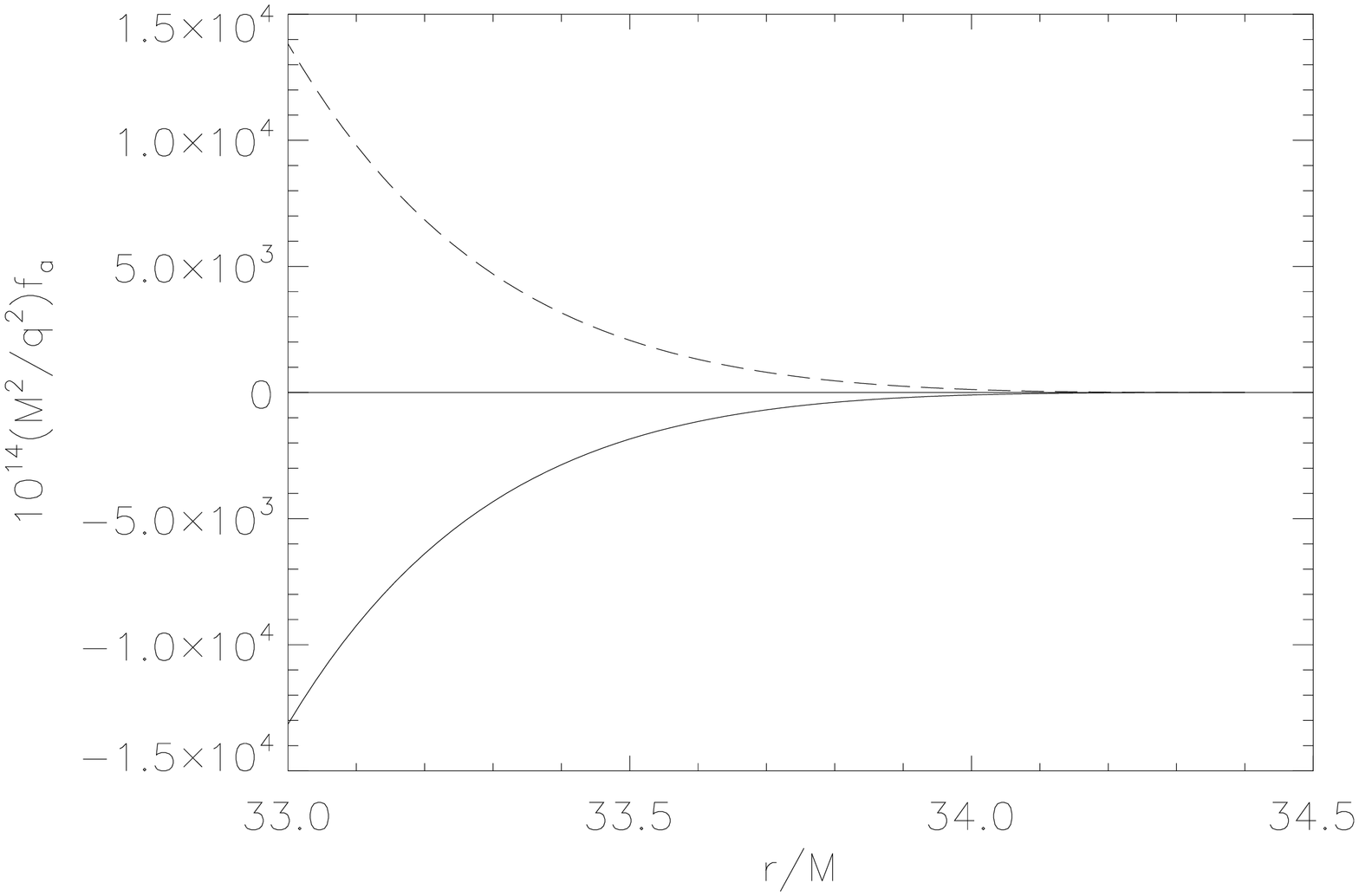}}}}
\end{picture}
\caption{In this figure the two components of the self-force are
plotted for the case of a particle undergoing radial infall after
being held at rest until a time $t=0$.  The particle's initial
location is at $r^* = 40M$ which corresponds to $r \approx 34.43 M$.
The dashed curve corresponds to $f_t$ and the solid curve to
$f_{r^*}$. } \label{plot}
\end{figure}
}}
 The expected convergence distance, obtained by measuring how far
the particle will fall in the time for a null geodesic leaving the
location of the particle at time $t=0$ to circle the black hole and
intersect the particle in a new location, is $r \approx 30.7 M$. On
the other hand, an examination of the error shows that more terms
than the six we have computed for each component of the self-force
are necessary for an accurate determination of the self-force when
$r < 33M$.

While the range over which we can compute the self-force is
relatively small and the calculation is for a particle held at rest
that subsequently undergoes radial infall in a Schwarzschild
spacetime, the merit of this method is it produces reliable results
and thus can provide an independent check on other (existing and
future) prescriptions for the calculation of the self-force. It is
worth noting that our calculation started with a finite result and
therefore no regularization was needed. The generalization to the
cases where the self-force is due to electromagnetism and gravity
should be straight-forward.

\acknowledgments
Some of this work was done while P.\ R.\ A.\ was visiting Los Alamos National
Laboratory.  He would like to thank E. Mottola and the entire T8 Group for their
hospitality.  This work was supported in part by the National
Science Foundation under grant PHY03-00710.

\appendix
\section{Calculation of  $u $ and $\sigma_{t_R}$}

In this appendix we calculate the quantities $u(y,y_R)$ and
$\sigma_{t_R}(y,y_R)$ with $y = (t,r,\theta_0,\phi_0)$ and $y_R =
(t_R,r_0,\theta_0,\phi_0)$; $t_R$ is the retarded time defined in
Eq.\ (\ref{eq:tR}).  The points $y$ and $y_R$ are separated by a
radial null geodesic.  From Eqs.\ (\ref{eq:sigma}) and
(\ref{eq:defu}) it is clear that
\begin{subequations}
\begin{eqnarray}
 u(y,y_R) &=& \left[ \frac{-det \left(-\sigma_{;\, \mu \nu'}(x,x') \right)}{\sqrt{-g(x)}
 \sqrt{-g(x')} } \right]_{ x \rightarrow y, \, x' \rightarrow
 y_R}^{1/2}  \label{eq:u1} \\
 \sigma_{t_R}(y,y_R) &=& \left[\frac{\partial}{\partial t'}
 \sigma(x,x') \right]_{ x \rightarrow y, \, x' \rightarrow
 y_R}  \;.
 \end{eqnarray}
 \end{subequations}

Although the points are separated by a radial null geodesic, it is
necessary to assume a more general separation before computing
derivatives.  Once this is done then the specific separation can be
taken.  Because of spherical symmetry the angular dependence of the
quantity $\sigma(x,x')$ must be such that it is a function of $\cos
\gamma$ which is defined in Eq.\ (\ref{eq:cosgamma}).  To see this
assume that the two points have an angular separation of $\gamma$.
Since any arbitrary rotation of the coordinate system does not
change this angle, the angular dependence of $\sigma$ can only be
through $\gamma$. Because any $2 \pi$ rotation around the origin of
the coordinate system should render $\sigma$ unchanged, $\sigma$
must be a periodic function of $\gamma$ with period $2 \pi$.  This
dependence can be written as a Fourier series in $\sin n \gamma$ and
$\cos n \gamma$. The coefficients of the sine terms must identically
vanish since $\sigma(x,x')$ is a symmetric function in $x$ and $x'$ and
switching $x$ and $x'$ changes the sign of $\gamma$.  Since $\cos n \gamma$
can be written in terms of powers of $\cos \gamma$~\cite{footnote}, it is then clear
that $\sigma =\sigma(t,r;t',r';\cos\gamma)$.

We next compute the angular derivatives of $\sigma$ beginning with
\begin{eqnarray}
\label{sigmatheta1}\sigma_{; \, \theta}(x,x') = \left(
\frac{\partial \sigma}{\partial \cos \gamma} \right)  \left( -\sin
\theta \cos \theta' + \cos \theta \sin \theta' \cos (\phi -
\phi')\right) \;.
\end{eqnarray}
Note that in general $\sigma_{; \, \mu}(x,x')$ is a vector at $x$
and a scalar at $x'$. Next we can proceed to take a second
derivative.
\begin{eqnarray}
\sigma_{; \, \theta \theta'}(x,x') =&\,& \left( \frac{\partial^2
\sigma}{\partial (\cos \gamma)^2} \right)  \left( -\sin \theta \cos
\theta' + \cos \theta \sin \theta' \cos (\phi - \phi')\right)^2
\nonumber \\  &\,& + \left( \frac{\partial \sigma}{\partial \cos
\gamma} \right)  \left( \sin \theta \sin \theta' + \cos \theta \cos
\theta' \cos (\phi - \phi')\right)
\end{eqnarray}
For the special case of a radial geodesic,  $\theta' = \theta$,
$\phi' = \phi$, and thus $\cos \gamma = 1$. As a result
\begin{subequations}
\label{sigmader}
\begin{eqnarray}
\label{sigmatheta} \left(\sigma_{; \, \theta \theta'} \right)_{\cos
\gamma = 1} = \left( \frac{\partial \sigma}{\partial \cos \gamma}
\right)_{\cos \gamma = 1} \;.
\end{eqnarray}
Similarly it is easy to show that
\begin{eqnarray}
\label{sigmaphi}\left(\sigma_{; \, \phi\phi'}\right)_{\cos \gamma =
1} = \sin^2 \theta \left(\frac{\partial \sigma}{\partial(\cos
\gamma)}\right)_{\cos \gamma = 1}
\end{eqnarray}
and
\begin{eqnarray}
\left(\sigma_{; \, \theta t'}\right)_{\cos \gamma =1} &=&
\left(\sigma_{; \, \theta r'}\right)_{\cos \gamma =1} =
  \left(\sigma_{; \, \theta \phi'}\right)_{\cos \gamma =1} = 0 \label{sigmatheta0}\\
\left(\sigma_{; \, \phi t'}\right)_{\cos \gamma =1} &=&
\left(\sigma_{; \, \phi r'}\right)_{\cos \gamma =1} =
  \left(\sigma_{; \, \phi \theta'}\right)_{\cos \gamma =1} = 0 \label{sigmaphi0} \;.
\end{eqnarray}
\end{subequations}
 Along with the other combinations of radial and
time derivatives it is still necessary to calculate $
\left(\frac{\partial \sigma}{\partial(\cos \gamma)} \right)_{\cos
\gamma = 1}$.  If we choose coordinates such that $\theta = \theta'
= \pi/2$ then from Eq.\ (\ref{sigmaphi})
\begin{equation}
\left(\frac{\partial \sigma}{\partial(\cos \gamma)} \right)_{\cos
\gamma = 1} = \left(\sigma_{; \, \phi \phi'}\right)_{\phi'=\phi} \;.
\end{equation}

To calculate the rest of the derivatives that comprise $\sigma_{; \,
\mu\nu'}$ it is useful to write $\sigma$ in terms of the proper time
$\tau$ and to assume that the points both lie in the equatorial
plane so that $\theta = \theta' = \pi/2$. Since $\sigma$ is one-half
the square of the proper distance between the two points along the
shortest geodesic connecting them we have
\begin{eqnarray}
\sigma &=& -\frac{1}{2}\tau^2  \label{sigma2} \;.
\end{eqnarray}
The proper time $\tau$ can be computed by solving the geodesic
equations~\cite{weinberg}
\begin{subequations}
\begin{eqnarray} \tau &=& -\sqrt{E}
\int_{r'}^{r} \frac{d\bar{r}}{ (1 - {E f(\bar{r}) - \frac{J^2
f(\bar{r})}{\bar{r}^2}})^{1/2}} \label{geodesic:tau} \\
\nonumber \\
  t - t' &=& - \int_{r'}^{r} \frac{d\bar{r}}{f(\bar{r})
(1 - E f(\bar{r}) -
\frac{J^2 f(\bar{r})}{\bar{r}^2})^{1/2}} \label{geodesic:t} \\ \nonumber \\
\nonumber  \\
\phi - \phi' &=& - J \int_{r'}^{r} \frac{d\bar{r}}{\bar{r}^2 (1 - E
f(\bar{r}) - \frac{J^2 f(\bar{r})}{\bar{r}^2})^{1/2}}
\label{geodesic:phi}
\end{eqnarray}
\label{eq:geodesic}
\end{subequations}
where $f(r) \equiv 1 - 2M/r$. Note that $J = 0$ yields a radial
geodesic and $E = 0$ a null one.

Expanding the integrand of Eq.\ (\ref{geodesic:tau}) in powers of
$E$ and $J$, computing the integral, and then substituting into Eq.\
(\ref{sigma2}) gives
\begin{equation}
\sigma = -\frac{E}{2} (r' - r)^2 - \frac{E^2}{2} F (r' - r) + O(E^3)
+ O(E J^2) \label{sigma3}
\end{equation}
with \begin{equation}
 F \equiv -\int_{r'}^r d \bar{r} f(\bar{r}) =
r' - r - 2M \ln \left(\frac{r'}{r}\right) \;. \label{eq:F}
\end{equation}
The values of the constants $E$ and $J$ can be found by solving
Eqs.\ (\ref{geodesic:t}) and (\ref{geodesic:phi}).  The integrals
can be computed by first expanding the integrands in powers of $E$
and $J$.  From Eq.\ (\ref{geodesic:phi}) one finds
\begin{equation}
J = \frac{r' r}{r' - r} \, (\phi - \phi') + O[E (\phi - \phi')] +
O[(\phi -\phi')^3] \;.  \label{eq:J}
\end{equation}
Keeping terms in the expansion of the integrand in Eq.\
(\ref{geodesic:t}) to $O(J^2)$ and $O(E^2)$ and using (\ref{eq:J})
yields
\begin{eqnarray}
E &=& -\frac{2 (t' - t_R )}{r' -r} - \frac{r' r}{(r' - r)^2} \,
(\phi - \phi')^2
     - \frac{3 (t' - t_R )^2 F}{(r'-r)^3} \nonumber \\
     & &   \; + O[(t' - t_R)^3] + O[(t' - t_R) (\phi-\phi')^2] +
     O[(\phi-\phi')^4]   \label{eq:E}
\end{eqnarray}
with
\begin{equation}
t_R = t + \int_{r'}^r \frac{d\bar{r}}{f(\bar{r})} = t - (r'-r) -
2M \ln \frac{r'-2M}{r-2M} \;. \label{eq:tR}
\end{equation}
Substituting into Eq.\ (\ref{sigma3}) gives
\begin{eqnarray}
\sigma &=& (t' - t_R) (r' -r) - \frac{F}{2 (r' -r)} \, (t' - t_R)^2
          + \frac{r' r}{2} \, (\phi - \phi')^2 \nonumber \\
   & & \;       + O[(t' - t_R)^3] + O[(t' - t_R) (\phi-\phi')^2] +
     O[(\phi-\phi')^4] \;.  \label{sigma4}
\end{eqnarray}

Computing the various derivatives of $\sigma$ and then setting $t' =
t_R$ and $\phi' = \phi$ gives
\begin{equation}
 \sigma_{t_R} = (r' -r)  \label{eq:sigmatR}
\end{equation}
and
\begin{eqnarray}
\sigma_{;\, t t'} &=& \frac{F}{r'-r} \nonumber \\
\sigma_{;\, tr'} &=& -1 + \frac{F}{f(r') (r'-r)}  \nonumber \\
\sigma_{;\, r t'} &=& -1 + \frac{F}{f(r) (r'-r)} \nonumber \\
\sigma_{;\, r r'} &=& - \frac{1}{f(r)} - \frac{1}{f(r')} +
\frac{F}{f(r) f(r') (r'-r)} \nonumber \\
\sigma_{; \, \phi \phi'} &=& - r r' \;. \label{sigmamunup}
\end{eqnarray}

Using Eqs.\ (\ref{sigmader}) and (\ref{sigmamunup}) it is easy to
show that
\begin{eqnarray}
\det [-\sigma_{; \, \mu \nu'}(y,y_R)] & = & \left(r r_0
\sin\theta_0\right)^2 \;.
\end{eqnarray}
Substituting into Eq.\ (\ref{eq:u1}) then gives
\begin{eqnarray}
u(y,y_R) = 1 \;.
\end{eqnarray}
This is an exact result.

\section{Power Series Expansion for the Self-Force}

In this appendix an expansion for the self-force in powers of $r -
r_0$ is derived.  The coefficients of the expansion depend on the
mass $M$ of the black hole, the radius $r_0$ at which the particle
begins falling, the radius $r$ which is its present location, and
the charge $q$.

To begin, consider $\Phi_s$ in either Eq.\ (\ref{eq:phis1}) or
(\ref{eq:phis2}).  Its contribution to the self-force is given in
Eq.\ (\ref{eq:selfforce1}).  From Eq.\ (\ref{eq:tR}) it is clear
that $\partial_t = \partial_{t_R}$.  Because $v(x,x_1)$ satisfies
the equation $\Box_x v = 0$ and because of the time translation and
time reversal invariance of the metric (\ref{eq:metric}), $v$ is a
function of $(t-t_1)^2$.  Making use of this fact one finds that
\begin{eqnarray}
\frac{\partial \Phi_s}{\partial t} &=& \frac{q}{8 \pi} \sqrt{1 -
\frac{2M}{r_0}} \left[ v(t,r;t_R,r_0) - \int_0^{t_R}
\frac{\partial}{\partial t_1} v(t,r;t_1,r_0) dt_1 \right]
\nonumber \\
&=& \frac{q}{8 \pi} \sqrt{1 - \frac{2M}{r_0}} v(t,r;0,r_0) \;.
\end{eqnarray}
Here as in Section \ref{sec:computation} we suppress the dependence
of $v$ on $\theta_0$ and $\phi_0$.

Next consider $\Phi_f$ in Eq.\ (\ref{eq:phif}).  The geodesic
equations (\ref{eq:geodesic}) can be used to change the integration
variable from the proper time to the coordinate time.  In this case
$J=0$ as the geodesic is radial. Since it starts from rest at $r_0$
it can be seen from the geodesic equations that $E = (1 -
2M/r_0)^{-1}$. The result is
\begin{equation}
\Phi_f = \frac{q}{8 \pi} \frac{1}{\sqrt{1-\frac{2M}{r_0}}} \int_0^t
v(t,r;t_1,r_1)
          \left(1 - \frac{2 M}{r_1}\right) \d t_1 \;.
\end{equation}
Noting that in a Schwarzschild spacetime $v(x,x) = 0$, one finds
\begin{eqnarray}
\frac{\partial \Phi_f}{\partial t} &=& -  \frac{q}{8 \pi}
\frac{1}{\sqrt{1-\frac{2M}{r_0}}} \int_0^t \left( \frac{\partial
v(t,r;t_1,r_1)}{\partial t_1} \right) \left(1 - \frac{2
M}{r_1}\right) d t_1 \;.
\end{eqnarray}
Since the particle is freely falling, $r_1 = r_1(t_1)$ and one can
write
\begin{eqnarray}
\frac{\partial v}{\partial t_1} \left(1 - \frac{2 M}{r_1}\right) &=&
\frac{d}{d t_1} \left[ v \left(1 - \frac{2 M}{r_1}\right) \right] -
\frac{d r_1}{d t_1} \frac{\partial}{\partial r_1} \left[v \left(1 -
\frac{2 M}{r_1}\right) \right]
\end{eqnarray}
with the result that
\begin{eqnarray}
\frac{\partial \Phi_f}{\partial t} &=& \frac{q}{8 \pi}
\frac{1}{\sqrt{1-\frac{2M}{r_0}}} \left\{ v[t,r;0,r_0] \left(1 -
\frac{2 M}{r_0}\right) \right.
\nonumber \\
& & \;\;\;\; \left. + \int_{r_0}^r \frac{\partial}{\partial r_1}
\left[ v(t,r;t_1,r_1) \left(1 - \frac{2 M}{r_1}\right) \right] d r_1
\right\} \;.
\end{eqnarray}
Thus
\begin{eqnarray}
f_t &=& - \frac{q^2}{8 \pi} \frac{1}{\sqrt{1-\frac{2M}{r_0}}}
\int_{r_0}^r \frac{\partial}{\partial r_1} \left[ v(t,r;t_1,r_1)
\left(1 - \frac{2 M}{r_1}\right) \right] d r_1
 \;.  \label{eq:ft1}
\end{eqnarray}

The computation of $f_{r*}$ is straightforward. Taking the
derivative of Eqs.\ (\ref{eq:phis2}) and (\ref{eq:phif}) and then
using the geodesic Eqs.\ (\ref{eq:geodesic}) and
(\ref{eq:rstar}) one finds
\begin{eqnarray}
f_{r*} &=& \frac{q^2}{8\pi}\sqrt{1 - \frac{2M}{r_0}} \left(1-\frac{2
M}{r} \right) \sum_{i,k=0}^\infty \left\{ \frac{1}{2i+1} \right. \nonumber \\
  & &  \left. \times \left[ \frac{d v_{i0k}}{d r} (r-r_0)^k + k v_{i0 k}
(r-r_0)^{k-1} \right]
 \left[ t^{2i+1} - (t - t_R)^{2i+1} \right] \right\}  \nonumber \\
  & &   + \frac{q^2}{8 \pi}
\left(1 - \frac{2M}{r}\right) \sqrt{\frac{r_0}{2M}}
      \int_{r_0}^r \frac{\partial v(t,r;t_1,r_1)}{\partial r} \sqrt{\frac{r_1}{r_0 - r_1}} d
      r_1
         \;.  \label{eq:fr1}
\end{eqnarray}

The next step is to expand $t$ in powers of $r_0-r$ and $t_1$ in
powers of $r_0-r_1$.  This is done using the geodesic equation
(\ref{geodesic:t}).   Then making the change of variables
\begin{eqnarray}
s &=& \sqrt{\frac{r_0 - r}{r_0}} \nonumber \\
s_1 &=& \sqrt{\frac{r_0 - r_1}{r_0}} \nonumber \\
x_0 &=& \sqrt{\frac{2 M}{r_0}} \nonumber \\
w_0 &=& \sqrt{1-\frac{2M}{r_0}} \label{eq:sxy}
\end{eqnarray}
gives
\begin{equation}
t = \frac{r_0}{x_0} \left[ \frac{2 s}{w_0}  +
 \left(\frac{2}{3} -  w^2_0 \right) \left(\frac{s}{w_0}\right)^3 + \; ... \;.\right]
\end{equation}
Substituting this and the corresponding expression for $t_1$ into
Eqs.\ (\ref{eq:ft1}) and (\ref{eq:fr1}), expanding in powers of $s$
and $s_1$, and computing the integrals gives
\begin{subequations}
\begin{eqnarray}
f_t &=& q^2 \sum_{n=3}^\infty \, a_{2n} s^{2n}  \\
f_{r*} &=& q^2 \sum_{n=2}^\infty \, b_{2n+1} s^{2n+1} \;.
\end{eqnarray}
Using a $ 16^{\rm th} $ order WKB expansion for $v(x,x')$ we find
\begin{eqnarray}
a_6 &=&    \frac{w_0\,\left( -13 + 53\,{w_0}^2 \right) }
   {22400\,\pi \,{r_0}^2}
 \nonumber \\
a_8 &=&      \frac{w_0}{470400\,\pi \,{r_0}^2\,
     {x_0}^2} \,\left( -703 + 2031\,{w_0}^2 +
       72\,{w_0}^4 \right)
  \nonumber \\
a_{10} &=&       \frac{w_0}{155232000\,\pi
\,{r_0}^2\,{x_0}^4}\,\left( -404385 + 1485889\,{w_0}^2 -
       1990813\,{w_0}^4 + 1497309\,{w_0}^6 \right)
 \nonumber \\
a_{12} &=&   \frac{w_0}{355170816000\,\pi \,
     {r_0}^2\,{x_0}^6} \,\left( -1380534784 + 6209685607\,{w_0}^2 -
9489117981\,{w_0}^4 \right. \nonumber \\
& & \; \left. + 2956643877\,{w_0}^6  +
       3255643281\,{w_0}^8 \right)
 \nonumber \\
a_{14} &=&    \frac{w_0}{129282177024000\,\pi \,
     {r_0}^2\,{x_0}^8}\,\left( -690105799163 +
       3725808468235\,{w_0}^2 \right. \nonumber \\
       & & \; \left.  -
       7969067122446\,{w_0}^4  +
       9871663249718\,{w_0}^6 -
       9633792321223\,{w_0}^8 \right. \nonumber \\
       & & \; \left. +
       5317042452879\,{w_0}^{10} \right)
\nonumber \\
a_{16} &=&    -\frac{w_0}{131867820564480000\,
     \pi \,{r_0}^2\,{x_0}^{10}} \left( 917523531344444 -
       5807610696686037\,{w_0}^2 \right. \nonumber \\
       & & \left. +
       15120303760167717\,{w_0}^4 -
       19548426682161946\,{w_0}^6 +
       7163353443413246\,{w_0}^8 \right. \nonumber \\
       & & \left. +
       11432898205974399\,{w_0}^{10} -
       9956772991427823\,{w_0}^{12} \right)
\end{eqnarray}
and
\begin{eqnarray}
b_5 &=&   -\frac{3\, x_0 \, {w_0}^2\,}{11200\,\pi \, r_0^2 \,
     }       \nonumber \\
b_7 &=&   \frac{1}{470400\, \pi \, x_0  r_0^2 }\left( -147 + 1031\,
{w_0}^2 - 2004\, {w_0}^4 \right)
   \nonumber \\
b_9 &=&   \frac{1}{558835200\,\pi \, x^3_0
     r_0^2}\left(725274 - 4840427\,{w_0}^2 + 7544064\,{w_0}^4 -
     1384911\,{w_0}^6 \right)
\nonumber \\
b_{11} &=&   \frac{1}{1598268672000\,\pi \,
     x_0^5 \, r_0^2} \left(-5192394350 + 35645887586\,{w_0}^2 -
     76724155827\,{w_0}^4  \right. \nonumber \\
     &  & \left.  + 74055537336\,{w_0}^6   -
     34933610745\,{w_0}^8 \right)
\nonumber \\
b_{13} &=&   \frac{1}{13851661824000\,\pi \,x_0^7 \, r_0^2}
     \left(88650418610 - 652839443586\,{w_0}^2 +
     1715470441205\,{w_0}^4  \right. \nonumber \\
     & & \left. - 2048237904519\,{w_0}^6    +
     938994638717\,{w_0}^8 + 27641613573\,{w_0}^{10} \right)
\nonumber \\
b_{15} &=&   \frac{1}{98900865423360000\,\pi \,
     x_0^9 \, r_0^2}
     \left( -1082712168000450 + 8690022729803252\,{w_0}^2  \right. \nonumber \\
     &  & \left.  -
     27039642594514215\,{w_0}^4  +
     43159798466443548\,{w_0}^6 -
     40035939562445204\,{w_0}^8  \right. \nonumber \\
     & & \left.  +
     24271024582691064\,{w_0}^{10}  -
     8499901220531595\,{w_0}^{12}\right) \;.
\end{eqnarray}
\end{subequations}

\section{Erratum}

In the Physical Review D (PRD) version of this paper~\cite{a-e-h} there were errors due to a factor of 2 error
in Paper I~\cite{AndHu}.  These errors have been corrected in this arXiv version and also published in an erratum~\cite{a-e-h-e}. For
completeness we list the corrections in that erratum here as well.

In Ref.~\cite{AndHu} there was a factor of two error in the calculation of the coefficients
of the expansion for $v(x,x')$ that has been corrected in Ref.~\cite{a-h-err}.
This expansion, carried to higher order, was used in the calculations of Ref. ~\cite{a-e-h}. We correct the
resulting errors in Ref.~\cite{a-e-h} here.

Eqs.\ (22.a) and (22.b) should be multiplied by a factor of 2 with the result
\begin{subequations}
\begin{eqnarray}
f_t(\tau) &=& \frac{q^2}{22400\,\pi \,
     r_0^2} \sqrt{1 - \frac{2M}{r_0}}
\left(40 - \frac{106M}{r_0} \right) \left(\frac{r_0 -
r}{r_0}\right)^3
     + O\left[\left(\frac{r_0-r}{r_0}\right)^4 \right] \\
f_{r*}(\tau) &=& - \frac{3 q^2}{11200 \pi r_0^2}\,
\sqrt{\frac{2M}{r_0}} \left(1 - \frac{2M}{r_0} \right)
\left(\frac{r_0-r}{r_0}\right)^{5/2} +
O\left[\left(\frac{r_0-r}{r_0} \right)^{7/2} \right]
\end{eqnarray}
\label{eq:fmu}
\end{subequations}

The entries in columns 2 and 4 of Table 1 should be multiplied by a factor of 2.  The revised results are shown in Table 1.
The curves in Figure 1 need to be multiplied by a factor of 2.  The revised figure is shown in Figure 1.

{\center{
\begin{table}
 \begin{tabular}{|c|c|c|c|c|}
      \hline
  $ \; r/M \; $ & $10^{14} (M^2/q^2) f_t$  &  Error & $10^{14} (M^2/q^2) f_{r^*}$ & Error \\ \hline
      $34.25$ & $ 6.93066 $ &  $ 6.9 \times 10^{-6}$ & $ -5.7384 $ &  $  3.4 \times 10^{-5}$ \\ \hline
      $34.00$ & $ 120.72$ &  $ 4.4 \times 10^{-4}$ & $ -96.78$ &  $1.4 \times 10^{-3}$ \\ \hline
      $33.75$ & $ 616$ &  $ 3.4 \times 10^{-3}$ & $ -520$ &  $ 8.4 \times 10^{-3}$ \\ \hline
      $33.50$ & $ 2080$ &  $ 0.013$ & $ -1840$ &  $  0.25$ \\ \hline
      $33.25$ & $ 5680$ &  $ 0.031$ & $ -5200$ &  $ 0.052$ \\ \hline
      $33.00$ & $ 13800$ &  $ 0.059 $ & $ -13200$ &  $  0.089$ \\ \hline
      $32.75$ & $ 30000$ &  $ 0.10$ & $ -30000$ &  $  0.13    $ \\ \hline
      $32.50$ & $ 66000  $ &  $ 0.14     $ & $ -62000  $ & $  0.18    $ \\ \hline
  \end{tabular}
  \caption{The dependence of the temporal ($f_t$) and radial ($f_{r^*}$)
components of the self-force on the radial distance $r/M$ are given
for a particle undergoing radial infall after being held at rest
until a time $t=0$.  The particle's initial location is at $r^* =
40M$ which corresponds to $r \approx 34.43 M$.
  The error shown for each case is an estimate of the relative error and is obtained
  by taking the absolute value of the
  ratio of the last term used in the series for the self-force to the entire series.}
  \label{restab}
  \end{table}
}}

 {\center{
\begin{figure}
\begin{picture}(200,250)(-200,0)
\put(-250,0)
{\rotatebox{90}{\resizebox{8cm}{!}{\includegraphics{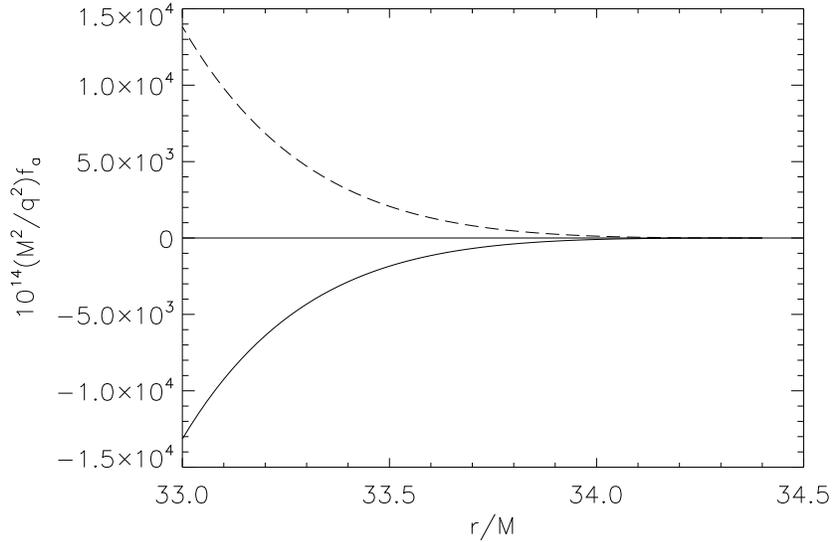}}}}
\end{picture}
\caption{In this figure the two components of the self-force are
plotted for the case of a particle undergoing radial infall after
being held at rest until a time $t=0$.  The particle's initial
location is at $r^* = 40M$ which corresponds to $r \approx 34.43 M$.
The dashed curve corresponds to $f_t$ and the solid curve to
$f_{r^*}$. } \label{plot}
\end{figure}
}}

Equations (B10c) and (B10d) need to be multiplied by a factor of 2.  The result is

\begin{subequations}
\begin{eqnarray}
a_6 &=&    \frac{w_0\,\left( -13 + 53\,{w_0}^2 \right) }
   {22400\,\pi \,{r_0}^2}
 \nonumber \\
a_8 &=&      \frac{w_0}{470400\,\pi \,{r_0}^2\,
     {x_0}^2} \,\left( -703 + 2031\,{w_0}^2 +
       72\,{w_0}^4 \right)
  \nonumber \\
a_{10} &=&       \frac{w_0}{155232000\,\pi
\,{r_0}^2\,{x_0}^4}\,\left( -404385 + 1485889\,{w_0}^2 -
       1990813\,{w_0}^4 + 1497309\,{w_0}^6 \right)
 \nonumber \\
a_{12} &=&   \frac{w_0}{355170816000\,\pi \,
     {r_0}^2\,{x_0}^6} \,\left( -1380534784 + 6209685607\,{w_0}^2 -
9489117981\,{w_0}^4 \right. \nonumber \\
& & \; \left. + 2956643877\,{w_0}^6  +
       3255643281\,{w_0}^8 \right)
 \nonumber \\
a_{14} &=&    \frac{w_0}{129282177024000\,\pi \,
     {r_0}^2\,{x_0}^8}\,\left( -690105799163 +
       3725808468235\,{w_0}^2 \right. \nonumber \\
       & & \; \left.  -
       7969067122446\,{w_0}^4  +
       9871663249718\,{w_0}^6 -
       9633792321223\,{w_0}^8 \right. \nonumber \\
       & & \; \left. +
       5317042452879\,{w_0}^{10} \right)
\nonumber \\
a_{16} &=&    -\frac{w_0}{131867820564480000\,
     \pi \,{r_0}^2\,{x_0}^{10}} \left( 917523531344444 -
       5807610696686037\,{w_0}^2 \right. \nonumber \\
       & & \left. +
       15120303760167717\,{w_0}^4 -
       19548426682161946\,{w_0}^6 +
       7163353443413246\,{w_0}^8 \right. \nonumber \\
       & & \left. +
       11432898205974399\,{w_0}^{10} -
       9956772991427823\,{w_0}^{12} \right)
\end{eqnarray}
and
\begin{eqnarray}
b_5 &=&   -\frac{3\, x_0 \, {w_0}^2\,}{11200\,\pi \, r_0^2 \,
     }       \nonumber \\
b_7 &=&   \frac{1}{470400\, \pi \, x_0  r_0^2 }\left( -147 + 1031\,
{w_0}^2 - 2004\, {w_0}^4 \right)
   \nonumber \\
b_9 &=&   \frac{1}{558835200\,\pi \, x^3_0
     r_0^2}\left(725274 - 4840427\,{w_0}^2 + 7544064\,{w_0}^4 -
     1384911\,{w_0}^6 \right)
\nonumber \\
b_{11} &=&   \frac{1}{1598268672000\,\pi \,
     x_0^5 \, r_0^2} \left(-5192394350 + 35645887586\,{w_0}^2 -
     76724155827\,{w_0}^4  \right. \nonumber \\
     &  & \left.  + 74055537336\,{w_0}^6   -
     34933610745\,{w_0}^8 \right)
\nonumber \\
b_{13} &=&   \frac{1}{13851661824000\,\pi \,x_0^7 \, r_0^2}
     \left(88650418610 - 652839443586\,{w_0}^2 +
     1715470441205\,{w_0}^4  \right. \nonumber \\
     & & \left. - 2048237904519\,{w_0}^6    +
     938994638717\,{w_0}^8 + 27641613573\,{w_0}^{10} \right)
\nonumber \\
b_{15} &=&   \frac{1}{98900865423360000\,\pi \,
     x_0^9 \, r_0^2}
     \left( -1082712168000450 + 8690022729803252\,{w_0}^2  \right. \nonumber \\
     &  & \left.  -
     27039642594514215\,{w_0}^4  +
     43159798466443548\,{w_0}^6 -
     40035939562445204\,{w_0}^8  \right. \nonumber \\
     & & \left.  +
     24271024582691064\,{w_0}^{10}  -
     8499901220531595\,{w_0}^{12}\right) \;.
\end{eqnarray}
\end{subequations}

We would like to thank Barry Wardell for alerting us to the
possibility that a factor of two error existed in our results in Ref.~\cite{a-h-err} and for help
in finding it.   This work was
supported in part by the National Science Foundation under grant
numbers PHY03-00710 and PHY05-56292.

\end{document}